\documentclass{elsarticle}
\usepackage{amsfonts}
\usepackage{amsmath}

\begin{document}

\begin{frontmatter}

\title{Non-autonomous H\'{e}non-Heiles system from Painlev\'{e} class}
\author{Maciej B\l aszak \\
Faculty of Physics, Division of Mathematical Physics, A. Mickiewicz
University \\
Umultowska 85, 61-614 Pozna\'{n}, Poland \\
blaszakm@amu.edu.pl}

\begin{abstract}
We show how to deform separable H\'{e}non-Heiles system with isospectral Lax representation, related with the stationary flow of the $5th$-order KdV,
to respective non-autonomous systems of Painlev\'{e} type with isomonodromic Lax representation.
\end{abstract}
\begin{keyword}
\textsl{Keywords:} non-autonomous Hamiltonian systems, Frobenius integrable
systems, isomonodromic Lax representation, Painlev\'{e} equations
\end{keyword}
\end{frontmatter}

There are two particular classes of second order nonlinear ordinary
differential equations (ODE's) playing important roles in modern physics and
mathematics. To the first class belong separable equations with autonomous
Hamiltonian representation. To the second class, belong Painlev\'{e}
equations with non-autonomous (in principle) Hamiltonian representation. The
separable equations can be expressed by so-called Lax representation in the
form of isospectral deformation equations while the Painlev\'{e} equations
can be expressed by Lax representation in the form of isomonodromic
deformation equations.

Actually, separable equations belong to the class of Liouville integrable
systems. A Liouville system on a $2n$-dimensional Poisson manifold $(M,\pi )$%
, where $\pi $ is a Poisson operator, is the set of dynamical equations of
the form%
\begin{equation}
\frac{\partial \xi }{\partial t_{r}}=X_{h_{r}}(\xi )=\pi dh_{r}\text{, \ \ \
}r=1,\ldots ,n  \label{0}
\end{equation}%
where $\xi \in M$ denotes points on $M$ and $h_{r}(\xi )$ are $n$
Poisson-commuting functions on $M$%
\begin{equation}
\left\{ h_{r},h_{s}\right\} _{\pi }:=\pi (dh_{r},dh_{s})=0\text{, \ \ \ }%
r,s=1,\ldots ,n  \label{kom0}
\end{equation}%
so that
\begin{equation}
\left[ X_{h_{r}},X_{h_{s}}\right] =0\text{ \ \ \ }r,s=1,\ldots ,n.
\label{kom1}
\end{equation}%
Since all the vector fields $X_{h_{r}\text{ }}$commute (\ref{kom1}), the
system (\ref{0}), as a Pfaffian system, has a common, unique (local)
solution $\xi (t_{1},\ldots ,t_{n},\xi _{0})$ through each point $\xi
_{0}\in M$ depending in general on all the evolution parameters $t_{k}$.
Further, let $L(\lambda ;\xi )$ and $U_{k}(\lambda ;\xi )$ \ be a matrices
that belong to some Lie algebra and which depend rationally on the
independent $\lambda $ called a spectral parameter. The autonomous separable
equations (\ref{0}) can be represented by the Lax form
\begin{equation}
\frac{\partial L(\lambda ;\xi )}{\partial t_{k}}=[U_{k}(\lambda ;\xi
),L(\lambda ;\xi )],  \label{8}
\end{equation}%
which is called the isospectral deformation equation because the eigenvalues
of the matrix $L$ are independent of all times $t_{k}$, $k=1,...,n.$

Now consider a set of $n$ non-autonomous Hamiltonians $H_{r}(\xi ,t)$
satisfying the Frobenius condition
\begin{equation}
\frac{\partial H_{r}}{\partial t_{s}}-\frac{\partial H_{s}}{\partial t_{r}}%
+\{H_{r},H_{s}\}=f_{rs}(t_{1},...,t_{n}),\quad r,s=1,\dots ,n  \label{fc}
\end{equation}%
instead of (\ref{kom0}) ones, where $f_{rs}$ are functions of evolution
parameters only. In consequence, the non-autonomous Hamiltonian vector fields

\begin{equation}
Y_{H_{k}}(\xi ,t)=\pi dH_{k}\text{, \ \ \ }k=1,\ldots ,n  \label{Y0}
\end{equation}%
satisfy the vector-field counterpart of (\ref{fc})
\begin{equation}
\frac{\partial Y_{H_{r}}}{\partial t_{s}}-\frac{\partial Y_{H_s}}{\partial
t_{r}}+\left[ Y_{H_{s}},Y_{H_{r}}\right] =0,\quad r,s=1,\dots ,n,
\label{fcv}
\end{equation}%
as $Y_{\{H_{r},H_{s}\}}=-\left[ Y_{H_{r}},Y_{H_{s}}\right] $. Therefore, the
set of non-autonomous Hamiltonian equations (the Pfaffian system)%
\begin{equation}
\frac{\partial \xi }{\partial t_{r}}=Y_{H_{r}}(\xi ,t)=\pi dH_{r}\text{, \ \
\ }r=1,\ldots ,n  \label{Pff}
\end{equation}%
has again common solutions $\xi (t_{1},\ldots ,t_{n},\xi _{0})$ through each
point $\xi _{0}$ of $M$ \cite{Fecko,J}.

If the non-autonomous Hamiltonian equations (\ref{Y0}) are of the Painlev%
\'{e} type then are represented by so-called Lax isomonodromic deformations.
This means that their solutions can be obtained from a system of linear
equations
\begin{equation}
\frac{\partial \Psi }{\partial \lambda }=L(\lambda ;\xi ,t)\Psi ,\qquad
\frac{\partial \Psi }{\partial t_{k}}=U_{k}(\lambda ;\xi ,t)\Psi ,
\label{11a}
\end{equation}%
where matrices $L$ and $U$ have rational singularities in $\lambda $, for
which the compatibility condition
\begin{equation}
\frac{\partial L(\lambda ;\xi ,t)}{\partial t_{k}}=[U(\lambda ;\xi
,t),L(\lambda ;\xi ,t)]+\frac{\partial U_{k}(\lambda ;\xi ,t)}{\partial
\lambda }  \label{11}
\end{equation}%
is equivalent to the corresponding Painlev\'{e} equation (\ref{Y0}). The
analytic continuation of a fundamental matrix solution for the first
equation in the system (\ref{11a}) defines monodromy data that is
independent of all $t_{k}$, what is ensured by the second equation, hence
the system (\ref{11}) is called an isomonodromy problem. Note also, that the
isomonodromy representation (\ref{11}) is only the necessary condition for
the Painlev\'{e} property \cite{D}, so equations with representation (\ref%
{11}) should be rather called of the Painlev\'{e} type.

The advantage of nonlinear separable ODE's is their integrability by
quadratures. As for Painlev\'{e} equations, although they are not integrable
by quadratures, nevertheless they have solutions which are free of movable
branch points and essential singularities. So, poles are the only
singularities of the solutions which change their position if one varies the
initial data. Thus, the solutions of the Painlev\'{e} ODE's are `regular'
single-valued functions around movable poles (meromorphic in the solution
domain), and as such are good candidates that define new special (\emph{%
transcendental}) functions.

A significant progress in construction of new multi-component Painlev\'{e}
equations took place since the modern theory of nonlinear integrable PDE's
has been born (the so-called \emph{soliton theory}). It was found that the
Painlev\'{e} equations are inseparably connected with the soliton systems
with whom they share many properties (see \cite{S1,S2,S3,S4,S5} and
references therein). The Painlev\'{e} equations are constructed under
particular reductions of soliton PDE's hierarchies.

In that short letter we would like to draw the attention of the reader onto
alternative way of construction of alredy known and new Painlev\'{e} type
ODE's by an appropriate deformations of separable ODE's. The method consists
of few steps. First, consider a separable geodesic motion on an appropriate $%
n $-dimensional pseudo-Riemannian space $(Q,g)$ with a metric $g$ that is
flat or of constant curvature. In Hamiltonian formalism on $M=T^{\ast }Q$,
with such system one can relates $n$ geodesic Hamiltonians $E_{1},...,E_{n}$
in involution and $n$ Hamiltonian vector fields $X_{1},...,X_{n}$ that
commute. Next, extend geodesic Hamiltonians $E_{i}\rightarrow \mathfrak{h}%
_{i}=E_{i}+W_{i},\ i=2,...,n$ by linear in momenta terms, generated by
Killing vectors of $g$ in such a way that $\mathfrak{h}_{i}$ constitute a
Lie algebra \cite{blasz2017}. Then, add separable potentials $\mathfrak{h}%
_{i}\rightarrow h_{i}=E_{i}+W_{i}+V_{i}$ and prove for which ones there
exists a non-autonomous deformation $h_{i}\rightarrow H_{i}(t_{1},...,t_{n})$
satisfying the Frobenius condition (\ref{fc}). The deformation procedure in
the geodesic case $\mathfrak{h}_{i}\rightarrow H_{i}(t_{1},...,t_{n})$ is
presented in \cite{BMS}. The systematic work on the deformation procedure
with nontrivial potentials is in progress. Finally, one should investigate
the related deformation of Lax representation, based on the results from
\cite{bl2018}.

Here, we would like to show the simple illustration of the method on the
example of one of the integrable cases of the celebrated H\'{e}non-Heiles
system and its deformation to non-autonomous system with isomonodromic Lax
representation. Slightly different deformation of that system, coming from
the similarity solutions of soliton equations was considered in \cite{Honey}.

Consider Liouville integrable extended H\'{e}non-Heiles system on $M=\mathbb{%
R}^{4},$ generated by two Hamiltonian functions
\begin{align}
h_{1}& =E_{1}+V_{1}(x)=\frac{1}{2}p_{1}^{2}+\frac{1}{2}p_{2}^{2}+x_{1}^{3}+%
\frac{1}{2}x_{1}x_{2}^{2}+\alpha \,x_{2}^{-2},  \notag \\
h_{2}& =E_{2}+V_{2}(x)=\frac{1}{2}x_{2}p_{1}p_{2}-\frac{1}{2}x_{1}p_{2}^{2}+%
\frac{1}{16}x_{2}^{4}+\frac{1}{4}x_{1}^{2}x_{2}^{2}-\alpha \,x_{1}x_{2}^{-2}
\label{hh1}
\end{align}%
in involution, written in Cartesian coordinates $(x_{1},x_{2})$ and
conjugate momenta $(p_{1},p_{2})$, where $E$ are geodesic parts of $h$,
while $V(x)$ are separable potentials. By setting the parameter $\alpha $
equal to zero we get one of the integrable cases of the standard H\'{e}%
non-Heiles system. The H\'{e}non-Heiles Hamiltonian is $h_{1}$, so for the
canonical form of the Poisson tensor $\{x_{i},p_{j}\}_{\pi }=\delta _{ij}$,
the related autonomous evolution equations are
\begin{align}
\frac{\partial x_{1}}{\partial t_{1}}& =\frac{\partial h_{1}}{\partial p_{1}}%
=p_{1},\ \ \ \ \ \frac{\partial x_{2}}{\partial t_{1}}=\frac{\partial h_{1}}{%
\partial p_{2}}=p_{2},  \notag \\
&  \label{3V1} \\
\frac{\partial p_{1}}{\partial t_{1}}& =-\frac{\partial h_{1}}{\partial x_{1}%
}=-3x_{1}^{2}-\frac{1}{2}x_{2}^{2},\ \ \ \ \ \frac{\partial p_{2}}{\partial
t_{1}}=-\frac{\partial h_{1}}{\partial x_{2}}=-x_{1}x_{2}+2\alpha
\,x_{2}^{-3}.  \notag
\end{align}%
What is important, equations (\ref{3V1}) represent the stationary flow of
the $5th$-order KdV \cite{Allan}. Here $h_{2}$ is the first integral of (\ref%
{3V1}) while the related equations
\begin{align}
\frac{\partial x_{1}}{\partial t_{2}}& =\frac{\partial h_{2}}{\partial p_{1}}%
=\frac{1}{2}x_{2}p_{2},\ \ \ \ \ \frac{\partial x_{2}}{\partial t_{2}}=\frac{%
\partial h_{2}}{\partial p_{2}}=\frac{1}{2}x_{2}p_{1}-x_{1}p_{2},  \notag \\
&  \label{3V2} \\
\frac{\partial p_{1}}{\partial t_{2}}& =-\frac{\partial h_{2}}{\partial x_{1}%
}=\frac{1}{2}p_{2}^{2}-\frac{1}{2}x_{1}x_{2}^{2}+\alpha \,x_{2}^{-2},\ \ \ \
\ \ \frac{\partial p_{2}}{\partial t_{2}}=-\frac{\partial h_{2}}{\partial
x_{2}}=-\frac{1}{2}p_{1}p_{2}-\frac{1}{4}x_{2}^{3}-\frac{1}{2}%
x_{1}^{2}x_{2}-2\alpha \,x_{1}x_{2}^{-3}  \notag
\end{align}%
represent the symmetry of (\ref{3V1}). Evolution equations (\ref{3V1}) and (%
\ref{3V2}) have Lax representations (\ref{8}), where \cite{bl2018}
\begin{gather*}
L(\lambda )=%
\begin{pmatrix}
p_{1}\lambda +\frac{1}{2}x_{2}p_{2} & \lambda ^{2}-x_{1}\lambda -\frac{1}{4}%
x_{2}^{2} \\
&  \\
-2\lambda ^{3}-2x_{1}\lambda ^{2}-\left( 2x_{1}^{2}+\frac{1}{2}%
x_{2}^{2}\right) \lambda +p_{2}^{2}+2\alpha \,x_{2}^{-2} & -p_{1}\lambda -%
\frac{1}{2}x_{2}p_{2}%
\end{pmatrix}%
, \\
\\
U_{1}(\lambda )=%
\begin{pmatrix}
0 & \frac{1}{2} \\
&  \\
-\lambda -2x_{1} & 0%
\end{pmatrix}%
,\quad U_{2}(\lambda )=%
\begin{pmatrix}
\frac{1}{2}p_{1} & \frac{1}{2}\lambda -\frac{1}{2}x_{1} \\
&  \\
-\lambda ^{2}-x_{1}\lambda -x_{1}^{2}-\frac{1}{2}x_{2}^{2} & -\frac{1}{2}%
p_{1}%
\end{pmatrix}%
.
\end{gather*}

Let us remark that for the geodesic Hamiltonians $E_{1}$ and $E_{2}$ there
exists infinite hierarchy of basic separable potentials, generated by the
recursion formula \cite{blasz1994,blasz2011}
\begin{equation}
V^{(k)}=\left(
\begin{array}{c}
V_{1}^{(k)} \\
V_{2}^{(k)}%
\end{array}%
\right) =R^{k}\left(
\begin{array}{c}
0 \\
1%
\end{array}%
\right) ,\ \ \ \ R=\left(
\begin{array}{cc}
x_{1} & 1 \\
\frac{1}{4}x_{2}^{2} & 0%
\end{array}%
\right) ,\ \ \ \ \ k\in \mathbb{Z}.  \label{rek}
\end{equation}%
The H\'{e}non-Heiles potential is the one for $k=4$ and the additional term
in (\ref{hh1}) is the potential with $k=-1$. The Lax representation for the
Hamiltonians with arbitrary linear combination of basic potentials the
reader can find in \cite{bl2018}.

Now, let us deform the original Hamiltonians (\ref{hh1}) in the following
way. First, subtract from $h_{2}$ the momentum $p_{1}$. Notice that $%
\{E_{1},p_{1}\}=0$, i.e. $W_{2}=-p_{1}$ is generated by the Killing vector $%
Z=(-1,0)^{T}$ of the Euclidean metric in $\mathbb{R}^{2}$. Second, add to
both Hamiltonians the lower nontrivial positive separable potentials (\ref%
{rek}) with coefficients depending on evolution parameters, i.e. $%
c_{3}(t_{1},t_{2})V^{(3)}+c_{2}(t_{1},t_{2})V^{(2)}$. Actually, consider the
following deformed Hamiltonians
\begin{align}
H_{1}(t)& =h_{1}+c_{3}(t_{1},t_{2})V_{1}^{(3)}+c_{2}(t_{1},t_{2})V_{1}^{(2)}
\notag \\
& =\frac{1}{2}p_{1}^{2}+\frac{1}{2}p_{2}^{2}+x_{1}^{3}+\frac{1}{2}%
x_{1}x_{2}^{2}+c_{3}(t_{1},t_{2})(x_{1}^{2}+\frac{1}{4}%
x_{2}^{2})+c_{2}(t_{1},t_{2})x_{1}+\alpha \,x_{2}^{-2},  \notag \\
H_{2}(t)&
=h_{1}-p_{1}+c_{3}(t_{1},t_{2})V_{2}^{(3)}+c_{2}(t_{1},t_{2})V_{2}^{(2)}
\label{dH} \\
& =\frac{1}{2}x_{2}p_{1}p_{2}-\frac{1}{2}x_{1}p_{2}^{2}-p_{1}+\frac{1}{16}%
x_{2}^{4}+\frac{1}{4}x_{1}^{2}x_{2}^{2}+\frac{1}{4}%
c_{3}(t_{1},t_{2})x_{1}x_{2}^{2}+\frac{1}{4}c_{2}(t_{1},t_{2})x_{2}^{2}-%
\alpha \,x_{1}x_{2}^{-2}.  \notag
\end{align}%
From the demand of the Frobenius condition (\ref{fc}) we immediately find
that
\begin{equation*}
c_{3}(t_{1},t_{2})=3t_{2},\ \ \ \ \ \ c_{2}(t_{1},t_{2})=t_{1}+3t_{2}^{2},\
\ \ \ f_{12}(t_{1},t_{2})=-c_{2}(t_{1},t_{2}).
\end{equation*}%
Hence, the related non-autonomous evolution equations are
\begin{align}
& \frac{\partial x_{1}}{\partial t_{1}}=\frac{\partial H_{1}}{\partial p_{1}}%
=p_{1},\ \ \ \ \ \ \ \ \frac{\partial x_{2}}{\partial t_{1}}=\frac{\partial
H_{1}}{\partial p_{2}}=p_{2},  \notag \\
&  \label{p1} \\
& \frac{\partial p_{1}}{\partial t_{1}}=-\frac{\partial H_{1}}{\partial x_{1}%
}=-3x_{1}^{2}-\frac{1}{2}x_{2}^{2}-6t_{2}x_{1}+t_{1}+3t_{2}^{2},\ \ \ \ \ \
\notag \\
&  \notag \\
& \frac{\partial p_{2}}{\partial t_{1}}=-\frac{\partial H_{1}}{\partial x_{2}%
}=-x_{1}x_{2}-\frac{3}{2}t_{2}x_{2}+2\alpha \,x_{2}^{-3}.  \notag
\end{align}%
and
\begin{align}
& \frac{\partial x_{1}}{\partial t_{2}}=\frac{\partial H_{2}}{\partial p_{1}}%
=\frac{1}{2}x_{2}p_{2}-1,\ \ \ \ \ \ \ \ \frac{\partial x_{2}}{\partial t_{2}%
}=\frac{\partial H_{2}}{\partial p_{2}}=\frac{1}{2}x_{2}p_{1}-x_{1}p_{2},
\notag \\
&  \notag \\
& \frac{\partial p_{1}}{\partial t_{2}}=-\frac{\partial H_{2}}{\partial x_{1}%
}=\frac{1}{2}p_{2}^{2}-\frac{1}{2}x_{1}x_{2}^{2}-\frac{3}{4}%
t_{2}x_{2}^{2}+\alpha \,x_{2}^{-2},  \label{p2} \\
&  \notag \\
& \frac{\partial p_{2}}{\partial t_{2}}=-\frac{\partial H_{2}}{\partial x_{2}%
}=-\frac{1}{2}p_{1}p_{2}-\frac{1}{4}x_{2}^{3}-\frac{1}{2}x_{1}^{2}x_{2}-%
\frac{3}{2}t_{2}x_{1}x_{2}-\frac{1}{2}\left( t_{1}+3t_{2}^{2}\right)
x_{2}-2\alpha \,x_{1}x_{2}^{-3}.  \notag
\end{align}%
The matrices $L(\lambda ,t),U_{1}(\lambda ,t)$ and $U_{2}(\lambda ,t)$ with
extra potential $3t_{2}V^{(3)}+(t_{1}+3t_{2}^{2})V^{(2)}$ are as follows
\cite{bl2018}
\begin{gather*}
L(\lambda ;t)=%
\begin{pmatrix}
p_{1}\lambda +\frac{1}{2}x_{2}p_{2} & \lambda ^{2}-x_{1}\lambda -\frac{1}{4}%
x_{2}^{2} \\
&  \\
-2\lambda ^{3}-2(x_{1}+3t_{2})\lambda ^{2}-\left( 2x_{1}^{2}+\frac{1}{2}%
x_{2}^{2}+6x_{1}t_{2}+6t_{2}^{2}+2t_{1}\right) \lambda +p_{2}^{2}+2\alpha
\,x_{2}^{-2} & -p_{1}\lambda -\frac{1}{2}x_{2}p_{2}%
\end{pmatrix}%
, \\
\\
U_{1}(\lambda ;t)=%
\begin{pmatrix}
0 & \frac{1}{2} \\
&  \\
-\lambda -2x_{1}-3t_{2} & 0%
\end{pmatrix}%
,\quad  \\
\\
U_{2}(\lambda ;t)=%
\begin{pmatrix}
\frac{1}{2}p_{1} & \frac{1}{2}\lambda -\frac{1}{2}x_{1} \\
&  \\
-\lambda ^{2}-(x_{1}+3t_{2})\lambda -x_{1}^{2}-\frac{1}{2}%
x_{2}^{2}-3x_{1}t_{2}-3t_{2}^{2}-t_{1} & -\frac{1}{2}p_{1}%
\end{pmatrix}%
.
\end{gather*}%
Now, because of explicit time dependence and the deformation of geodesic
Hamiltonian $E_{2}$ by $W_{2}=-p_{1}$ term, we get
\begin{eqnarray*}
\frac{\partial L(\lambda ;t)}{\partial t_{1}}-[U_{1}(\lambda ;t),L(\lambda
;t)] &=&\left(
\begin{array}{cc}
0 & 0 \\
-2\lambda  & 0%
\end{array}%
\right) =2\lambda \frac{\partial U_{1}(\lambda ;t)}{\partial \lambda }, \\
&& \\
\frac{\partial L(\lambda ;t)}{\partial t_{2}}-[U_{2}(\lambda ;t),L(\lambda
;t)] &=&\left(
\begin{array}{cc}
0 & \lambda  \\
-4\lambda ^{2}-2(x_{1}+3t_{2})\lambda  & 0%
\end{array}%
\right) =2\lambda \frac{\partial U_{2}(\lambda ;t)}{\partial \lambda }
\end{eqnarray*}%
and so, the non-autonomous evolution equations (\ref{p1}) and (\ref{p2})
have the following isomonodromic Lax representation%
\begin{equation*}
\frac{\partial L(\lambda ;t)}{\partial t_{k}}=[U(\lambda ;t),L(\lambda
;t)]+2\lambda \frac{\partial U_{k}(\lambda ;t)}{\partial \lambda },\ \ \ \ \
k=1,2,
\end{equation*}%
or the (\ref{11}) one after reparametrization of spectral parameter $\lambda
\rightarrow \exp (2\lambda )$.

The presented non-autonomous system seems to belong to the $P_{II}$%
-hierarchy as the extended H\'{e}non-Heiles evolution equations (\ref{3V1})
represent the stationary flow of the $5th$-order KdV, but we could not find
in the literature neither the system (\ref{dH}) nor its isomonodromy
representation in explicit form.

The complete theory of such deformations, with many other examples and the
classification of hierarchies, will be presented in subsequent articles.

\end{document}